\documentclass[a4paper,conference]{IEEEtran}

\usepackage{ifthen}
\usepackage{amsfonts}
\usepackage{epsfig}
\usepackage{float}
\usepackage{color}
\usepackage{multicol}
\usepackage{mathcomste}
\usepackage{amsthm}
\usepackage{multicol}
\usepackage{subfigure}
\usepackage{cite}
\usepackage{times}

\usepackage{amsmath,amssymb,amsfonts,eucal,latexsym}
\usepackage{graphicx}
\usepackage{pst-plot}
\usepackage{pstricks}
\usepackage{color}

\newtheorem{thm}{Theorem}

\newcommand{\reff}[1]{\eqref{#1}}

\def \gapTot {1.87~}

\begin{document}
\title{State of the cognitive interference channel: a new unified inner bound}

\author{%
\IEEEauthorblockN{Stefano Rini, Daniela Tuninetti and Natasha Devroye}
\IEEEauthorblockA{University of Illinois at Chicago\\
             Chicago, IL 60607, USA\\
              Email: srini2, danielat, devroye@uic.edu}
}
%




\maketitle

\begin{abstract}
The capacity region of the interference channel in which one transmitter non-causally knows the message of the other, termed the cognitive interference channel, has remained open since its inception in 2005. A number of subtly differing achievable rate regions and outer bounds have been derived, some of which are tight under specific conditions. In this work we present a new unified inner bound for the discrete memoryless cognitive interference channel. We show explicitly how it   encompasses all known discrete memoryless achievable rate regions as special cases. The presented achievable region was recently used in deriving the capacity region of the linear high-SNR deterministic approximation of the Gaussian cognitive interference channel. The high-SNR deterministic approximation was then used to obtain the capacity of the Gaussian cognitive interference channel to within \gapTot bits.
 \end{abstract}

\section{Introduction}


The cognitive interference channel (CIFC)\footnote{Other names for this channel include the cognitive radio channel \cite{devroye_IEEE}, interference channel with degraded message sets \cite{JiangXinAchievableRateRegionCIFC, WuDegradedMessageSet}, the non-causal interference channel with one cognitive transmitter \cite{biao2009}, the interference channel with one cooperating transmitter \cite{MaricGoldsmithKramerShamai07Eu} and the interference channel with unidirectional cooperation \cite{MaricUnidirectionalCooperation06, maric2005capacity}.} is an interference channel in which one of the transmitters - dubbed the cognitive transmitter - has non-causal knowledge of the message of the other - dubbed the primary - transmitter. The study of this channel is motivated by cognitive radio technology which allows wireless devices to sense and adapt to their RF environment by changing their transmission parameters in software on the fly. One of the driving applications of cognitive radio technology is secondary spectrum sharing: currently licensed spectrum would be shared by  primary (legacy) and secondary (usually cognitive) devices in the hope of improving spectral efficiency. The extra abilities of cognitive radios may be modeled information theoretically in a number of ways - see \cite{goldsmith_survey, devroye_chapter1} for surveys -  one of which is through the assumption of non-causal primary message knowledge at the secondary, or cognitive, transmitter.


The two-dimensional capacity region of the CIFC has remained open in general since its inception in 2005 \cite{devroye_CISS2005}. However, capacity is known in a number of classes of channels:

\smallskip

\noindent $\bullet$ {\bf General deterministic CIFCs.} The capacity region of  fully deterministic CIFCs in the flavor of the deterministic interference channel \cite{elgamal_det_IC} has been obtained in \cite{Rini-prelim}. 
A special case of the deterministic CIFC is the deterministic  linear high-SNR approximation of the Gaussian CIFC, whose capacity region,  in the spirit of \cite{Avestimehr:2007:ITW}, was obtained in \cite{usItw09}. \\
$\bullet$  {\bf Semi-deterministic CIFCs.} In \cite{biao2009} the capacity region for a class of  channels in which the signal at the cognitive receiver is a deterministic function of the channel inputs is derived. \\
$\bullet$  {\bf Discrete memoryless CIFCs.} First considered in \cite{devroye_CISS2005, devroye_IEEE}, its capacity region was obtained for very strong interference in \cite{MaricUnidirectionalCooperation06} and for weak interference in \cite{WuDegradedMessageSet}. Prior to this work and the recent work of \cite{biao2009}, the largest known achievable rate regions were those of \cite{MaricGoldsmithKramerShamai07Eu, devroye_IEEE, DevroyeThesis, JiangXinAchievableRateRegionCIFC}. The recent and independently derived region of \cite{biao2009} was shown to contain \cite{MaricGoldsmithKramerShamai07Eu, JiangXinAchievableRateRegionCIFC}, but was not conclusively shown to encompass \cite{devroye_IEEE} or the larger region of \cite{DevroyeThesis}. \\
$\bullet$ {\bf Gaussian CIFC.} The capacity region under weak interference was obtained in \cite{JovicicViswanath06, WuDegradedMessageSet}, while that for very strong interference follows from \cite{MaricUnidirectionalCooperation06}. Capacity for a class of Gaussion MIMO CIFCs is obtained in \cite{SridharanVishwanathMIMOCognitive}. \\
$\bullet$ {\bf Z-CIFCs.} Inner and outer bounds  when the cognitive-primary link is noiseless are obtained in \cite{Maric_Zchannnel_08, biao2008}. The Gaussian causal case is considered in  \cite{biao2009}, and is related to the general (non Z) causal CIFC explored in \cite{seyedmehdi2009}. \\
$\bullet$ {\bf CIFCs with secrecy constraints.} Capacity of a CIFC in which the cognitive message is to be kept secret from the primary and the cognitive wishes to decode both messages is obtained in \cite{liang-cognitive}. A cognitive multiple-access wiretap channel is considered in \cite{simeone_CISS}.

We focus on the discrete memoryless CIFC (DM-CIFC) and propose a new achievable rate region which encompasses all  other known achievable rate regions. 
We will explicitly demonstrate how our new region encompasses and may be reduced to the other regions. 
The new unified achievable rate region has been shown to be useful as:
 1) specific choices of random variables yield capacity in  the deterministic CIFC \cite{Rini-prelim} and hence also in the 2)    linear high-SNR approximation of the Gaussian CIFC \cite{usItw09},
  3) specific choices of Gaussian random variables have resulted in an achievable rate region which lies within \gapTot bits, 
  regardless of channel parameters, of an outer bound \cite{usITW10}. Numerical simulations indicate the actual gap is smaller.



\section{Channel Model}
\label{sec:channelmodel}

\begin{figure}[h]
  \centering
  \epsfig{figure=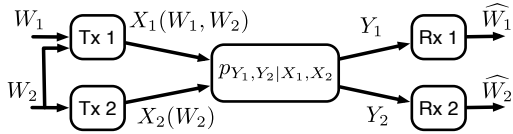, width=7cm}
  \caption{The Cognitive Interference Channel.}
  \label{fig:generalCIFC}
\end{figure}

The Discrete Memoryless Cognitive InterFerence Channel (DM-CIFC), as shown in Fig. \ref{fig:generalCIFC},  consists of two transmitter-receiver pairs
that exchange independent messages over a common channel. Transmitter $i$, $i\in\{1,2\}$, has discrete input alphabet $\Xcal_i$ and its receiver has discrete output alphabet $\Ycal_i$. The channel is assumed to be memoryless with transition probability $p_{Y_1,Y_2|X_1,X_2}$.
Encoder $i$, $i\in \{1,2\}$,  wishes to communicate a message $W_i$ uniformly distributed on
${\cal M}_i = [1: 2^{N R_i}]$ to decoder $i$ in $N$ channel uses at rate $R_i$.
Encoder~1 (i.e., the cognitive user) knows its own message $W_1$ and that of encoder~2 (the primary user), $W_2$. A rate pair $(R_1,R_2)$ is achievable if there exist
sequences of encoding functions
\begin{align*}
X_1^N &=  f_1^N(W_1, W_2), \;\;  f_1: {\cal M}_1 \times {\cal M}_2 \rightarrow {\cal X}_1^N, \\
X_2^N &=  f_2^N(W_2), \;\; \;\;\;\;\;\; f_2: {\cal M}_2 \rightarrow {\cal X}_2^N,
\end{align*}
with corresponding sequences of decoding  functions
\begin{align*}
\widehat{W}_1 & =  g_1^N(Y_1^N), \;\; g_1: {\cal Y}_1^N \rightarrow {\cal M}_1, \\
\widehat{W}_2 & =  g_2^N(Y_2^N),   \;\; g_2: {\cal Y}_2^N \rightarrow {\cal M}_2.
\end{align*}
The capacity region is defined as the closure of the region of achievable $(R_1,R_2)$ pairs
 \cite{ThomasCoverBook}.
Standard strong-typicality is assumed; properties may be found in
 \cite{kramerBook}.

\section{A new unified achievable rate region}
\label{sec:newregion}

As the DM-CIFC encompasses classical interference, multiple-access and broadcast channels, we expect to see a combination of their achievability proving techniques surface in any unified scheme for the CIFC: 

\smallskip

\noindent $\bullet$ {\bf Rate-splitting.} As in Han and Kobayashi \cite{Han_Kobayashi81} for the interference-channel and in the DM-CIFC regions of \cite{MaricGoldsmithKramerShamai07Eu, devroye_IEEE, JiangXinAchievableRateRegionCIFC}, rate-splitting is not necessary in the weak \cite{WuDegradedMessageSet} and strong \cite{MaricUnidirectionalCooperation06} interference regimes. \\
$\bullet$  {\bf Superposition-coding.} Useful in multiple-access and broadcast channels \cite{ThomasCoverBook}, in the CIFC the superposition of  private messages on top of common ones \cite{MaricGoldsmithKramerShamai07Eu, JiangXinAchievableRateRegionCIFC} is proposed and is known to be capacity achieving in very strong interference \cite{MaricUnidirectionalCooperation06}. \\
$\bullet$  {\bf Binning.} Gel'fand-Pinsker coding \cite{GelFandPinskerClassic}, often referred to as binning, allows a transmitter to "cancel" (portions of) the interference known to it at its intended receiver.  Related binning techniques are used by Marton in deriving the largest known DM-broadcast channel achievable rate region \cite{MartonBroadcastChannel}.

\begin{figure*}
\epsfig{figure=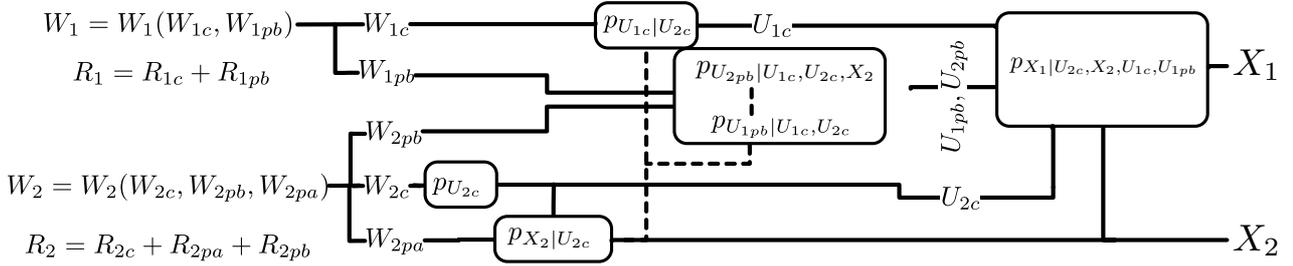, width=17cm}
 \caption{The achievable encoding scheme of Thm \ref{thm:our achievable region}. The ordering from left to right and the distributions demonstrate the codebook generation process. The dotted lines indicate binning. We see rate splits are used at both users, private messages $W_{1pb}, W_{2pa}, W_{2pb}$ are superimposed on common messages $W_{1c}, W_{2c}$ and $U_{1c}$ is binned against $X_2$ conditioned on $U_{2c}$, while $U_{1pb}$ and $U_{2pb}$ are binned against each and $X_2$  in a Marton-like fashion (conditioned on other subsets of random variables).}
 \label{fig:encodingScheme}
\end{figure*}

We now present a new achievable region for the DM-CIFC which
generalizes all best known achievable rate regions including
\cite{MaricGoldsmithKramerShamai07Eu,WuDegradedMessageSet, JiangXinAchievableRateRegionCIFC, devroye_IEEE} as well as \cite{biao2009}.  
\begin{thm}
\label{thm:our achievable region} {Region ${\cal R}_{RTD}$.}
A rate pair $(R_1,R_2)$ such that
\ea{
R_1 &=& R_{1c}+R_{1pb},\\
R_2 &=& R_{2c}+R_{2pa}+R_{2pb}
}
is achievable for a DM-CIFC if $(R_{1c}',R_{1pb}',R_{2pb}',R_{1c},$ $R_{1pb},R_{2c},R_{2pa},R_{2pb})\in\RR^8_+$ satisfies \eqref{eq:our achievable region R0'}--\eqref{eq:our achievable region Ed1:3}
\begin{figure*}
\begin{subequations}
\ea{
R{'}_{1c}                        &\geq& I(U_{1c};X_{2}|U_{2c})
\label{eq:our achievable region R0'}
\\
R{'}_{1c}+R{'}_{1pb}             &\geq& I(U_{1pb}, U_{1c}; X_{2}|U_{2c})
\label{eq:our achievable region R0'+R1'}
\\
R{'}_{1c}+R{'}_{1pb}+R{'}_{2pb} &\geq&  I(U_{1pb}, U_{1c}; X_{2} |U_{2c})+I(U_{2pb};  U_{1pb}| U_{1c},U_{2c},X_2)
\label{eq:our achievable region R1'+R2'}
\\
R_{2c}+R_{2pa}+(R_{1c}+R{'}_{1c})+(R_{2pb}+R{'}_{2pb}) &\leq& I(Y_2; U_{2pb},U_{1c},X_{2},U_{2c}) +I(U_{1c}; X_{2}| U_{2c})
\label{eq:our achievable region Ed2:1}
\\
       R_{2pa}+(R_{1c}+R{'}_{1c})+(R_{2pb}+R{'}_{2pb}) &\leq& I(Y_2; U_{2pb},U_{1c},X_{2}| U_{2c}) +I(U_{1c}; X_{2}| U_{2c})
\label{eq:our achievable region Ed2:2a}
\\
       R_{2pa}                   +(R_{2pb}+R{'}_{2pb}) &\leq& I(Y_2; U_{2pb},X_{2}| U_{1c},U_{2c}) +I(U_{1c}; X_{2}| U_{2c})
\label{eq:our achievable region Ed2:2b}
\\
               (R_{1c}+R{'}_{1c})+(R_{2pb}+R{'}_{2pb}) &\leq& I(Y_2; U_{2pb}, U_{1c}| X_{2},U_{2c})+I(U_{1c}; X_{2}| U_{2c})
\label{eq:our achievable region Ed2:3a}
\\
                                  (R_{2pb}+R{'}_{2pb}) &\leq& I(Y_2; U_{2pb}| U_{1c},X_{2},U_{2c})
\label{eq:our achievable region Ed2:3b}
\\
R_{2c}+(R_{1c}+R{'}_{1c})+(R_{1pb}+R{'}_{1pb}) &\leq& I(Y_1; U_{1pb},U_{1c},U_{2c}),
\label{eq:our achievable region Ed1:1}
\\
       (R_{1c}+R{'}_{1c})+(R_{1pb}+R{'}_{1pb}) &\leq& I(Y_1; U_{1pb},U_{1c}|U_{2c}),
\label{eq:our achievable region Ed1:2}
\\
                          (R_{1pb}+R{'}_{1pb}) &\leq& I(Y_1; U_{1pb}|U_{1c},U_{2c}),
\label{eq:our achievable region Ed1:3}
}
\label{eq:our achievable region}
\end{subequations}
\end{figure*}
for some input distribution
\[
p_{U_{1c},U_{2c},U_{1pb},U_{2pb}}
 p_{X_1,X_2|U_{1c},U_{2c},U_{1pb},U_{2pb}}
 p_{Y_1,Y_2|X_1,X_2}.
\]
\end{thm}
The encoding scheme used in deriving this achievable rate region is shown in Fig.\ref{fig:encodingScheme}.
The key aspects of our scheme are the following, where we drop $n$ for convenience:

$\bullet$
We {\bf rate-split} the independent messages $W_1$ and $W_2$ uniformly distributed on ${\cal M}_1 = [1:2^{nR_1}]$  and ${\cal M}_2 = [1:2^{nR_2}]$  into the messages $W_{i}$, $ i\in\{1c,2c,1pb,2pb,2pa\}$,  all independent and
uniformly distributed on $[1:2^{n R_i}]$, each encoded using the random variable $U_i$, such that
\begin{align*}
W_1 &=(W_{1c},W_{1pb}),  \;\;\;\;\;\;\; \;\;\;\;  \;R_1=R_{1c}+R_{1pb}, \\
W_2 &=(W_{2c},W_{2pb},W_{2pa}), \;\; R_2=R_{2c}+R_{2pa}+R_{2pb}.
\end{align*}


$\bullet$ {\bf Tx2 (primary Tx):} 
Transmitter~2 sends $X_2$ that carries the private message $W_{2pa}$ (``p'' for private, ``a'' for alone)  {\bf{superimposed}} to the common message $W_{2c}$ carried by $U_{2c}$ (``c'' for common).

 $\bullet$ {\bf Tx1 (cognitive Tx):} 
The common message of Tx1, encoded by $U_{1c}$, is {\bf binned}
against $X_2$ conditioned on $U_{2c}$.
The private message of Tx2, $W_{2pb}$, encoded by $U_{2pb}$ (``b'' for broadcast)
and a portion of the private message of Tx1, $W_{1pb}$, encoded as $U_{1pb}$, are {\bf binned} against each other as
in Marton's region \cite{MartonBroadcastChannel} conditioned on
$U_{1c}, U_{2c}$ and $U_{1c},U_{2c},X_2$ respectively.
%

Tx1 sends $X_1$ over the channel.
The incorporation of a Marton-like scheme at the cognitive transmitter was initially motivated by the fact that in certain regimes, this strategy was shown to be capacity achieving for the linear high-SNR deterministic CIFC \cite{usItw09}. \\


The codebook generation, encoding and decoding as well as the error event analysis is provided in \cite{Rini-prelim}. 

{\it Remark:}

\noindent $\bullet$ \reff{eq:our achievable region Ed2:1} can be dropped when $R_{2c}=R_{2pa}=R_{2pb}=R_{2pb}'=0$

\noindent $\bullet$ \reff{eq:our achievable region Ed2:2a} can be dropped when $R_{2pa}=R_{2pb}=R_{2pb}'=0$

\noindent $\bullet$ \reff{eq:our achievable region Ed2:3a} can be dropped when $R_{2pb}=R_{2pb}'=0$
  
\noindent $\bullet$ \reff{eq:our achievable region Ed1:1} can be dropped when $R_{1c}=R_{1c}'=R_{1pb}=R_{1pb}'=0$

\section{Comparison with existing achievable regions}
\label{sec:ComparisonNewAchievableRegion}

We now show that the  region of Theorem \ref{thm:our achievable region} contains all other known achievable rate
regions for the DM-CIFC.  We note that showing inclusion of the rate regions \cite[Thm.2]{biao2009}, \cite{jiang-achievable-BCCR}, and  \cite{DevroyeThesis} is sufficient to demonstrate the largest known DM-CIFC region, since the region of  \cite{biao2009} is shown to contain those of \cite[Th.1]{MaricGoldsmithKramerShamai07Eu} and  \cite{JiangXinAchievableRateRegionCIFC}, and the region of \cite{jiang-achievable-BCCR} is claimed to contain all others. The region in \cite{DevroyeThesis} is explicitly shown, for the first time, to be included in another region.    

\subsection{Devroye et al.'s region  \cite[Thm. 1]{DevroyeThesis}}
In the appendix we show that the region of  \cite[Thm. 1]{DevroyeThesis} ${\cal R}_{DMT}$, is contained in our new region  ${\cal R}_{RTD}$ along the lines:

\noindent
$\bullet$ We make a correspondence between the random variables and corresponding rates of ${\cal R}_{DMT}$ and ${\cal R}_{RTD}$.  \\
$\bullet$ We define new regions  $\Rcal_{DMT}\subseteq \Rcal_{DMT}^{out}$ and $\Rcal_{RTD}^{in} \subseteq \Rcal_{RTD}$ which are easier to compare: they have  identical input distribution decompositions and similar rate equations. \\
$\bullet$ For any fixed input distribution, an equation-by-equation comparison leads to  $\Rcal_{DMT} \subseteq \Rcal_{DMT}^{out} \subseteq  \Rcal_{RTD}^{in} \subseteq  \Rcal_{RTD}$.

\subsection{Cao and Chen's region \cite[Thm. 2]{biao2009}}
\label{subsec:biao}
The independently derived region in \cite[Thm. 2]{biao2009} uses a similar encoding structure as that of $\Rcal_{RTD}$ with two exceptions: a) the binning is done sequentially rather than jointly as in $\Rcal_{RTD}$ leading to binning constraints (43)--(45) in \cite[Thm. 2]{biao2009} as opposed to \eqref{eq:our achievable region R0'}--\eqref{eq:our achievable region R1'+R2'} in Thm.\ref{thm:our achievable region}. Notable is that both schemes have adopted a Marton-like binning scheme at the cognitive transmitter, as first introduced in the context of the CIFC in \cite{biao2008}. b) While the cognitive messages are rate-split in identical fashions, the primary message is split into 2 parts in \cite[Thm. 2]{biao2009} ($R_1=R_{11}+R_{10}$, note the reversal of indices) while we explicitly split the primary message into three parts $R_2 = R_{2c}+R_{2pa}+R_{2pb}$. In the Appendix we show that the region of \cite[Thm.2]{biao2009}, denoted as ${\cal R}_{CC} \subseteq {\cal R}_{RTD}$ in  two steps:

\noindent $\bullet$  We first show that we may WLOG set $U_{11} = \emptyset$ in \cite[Thm.2]{biao2009}, creating a new region $R_{CC}'$.

\noindent $\bullet$  We next make a correspondence between our random variables and those of \cite[Thm.2]{biao2009} and  obtain identical regions. 

\subsection{Jiang et al.'s region \cite[Thm. 4.1]{jiang-achievable-BCCR}}

The scheme originally designed for the more general broadcast channel with cognitive relays (or interference-chanel with a cognitive relay) may be tailored/reduced to derive a region for the cognitive interference channel.  This scheme also incorporates a broadcasting strategy. However, the common messages are created independently instead of having the common message from transmitter~1 being superposed to the common message from transmitter~2. The former choice introduces more rate constraints than the latter and allows us to show inclusion in $\Rcal_{RTD}$ after equating random variables.





\section{Conclusion}
A new achievable rate region for the DM-CIFC has been derived and shown to encompass all known achievable rate regions. Of note is the inclusion of a Marton-like broadcasting scheme at the cognitive transmitter. Specific choices of this region have been shown to achieve capacity for the linear high-SNR approximation of the Gaussian CIFC \cite{usItw09, Rini-prelim}, and the deterministic  CIFC in general \cite{Rini-prelim}. This region has furthermore been shown to achieve within \gapTot bits of an outer bound, regardless of channel parameters in \cite{usITW10, Rini-prelim}. Numerical evaluation of the region under Gaussian input distributions for the Gaussian CIFC is currently underway, while extensions of the CIFC to multiple users will be investigated in the longer term.


\newpage

\onecolumn

\begin{appendix}

\subsection{Proof that $X_{2a} = \emptyset$ WLOG in \cite[Th.1]{MaricGoldsmithKramerShamai07Eu}}
In their notation, after the Fourier-Motzkin elimination  of ~\cite[Th.1]{MaricGoldsmithKramerShamai07Eu} we obtain the achievable rate region
\begin{align}
R_1 &\leq I(U_{1a}; Y_1| U_{1c} , Q)-I(U_{1a}; X_{2a}, X_{2b}| U_{1c}, Q) \notag \\
& \;\;\;\;\;\; +I(X_{2b}, U_{1c}; Y_2 | X_{2a}, Q)
\label{eq:maric scheme: R1 1}\\
R_1 &\leq I(U_{1a},U_{1c} ; Y_1| Q)-I(U_{1a},U_{1c}; X_{2a}, X_{2b}|  Q)
\notag \\ 
R_2 &\leq I(X_2, U_{1c}; Y_2|Q)
\notag \\ 
R_2 &\leq I(X_2; Y_2, U_{1c}|Q)
\notag \\ 
R_1+R_2 &\leq I(U_{1a}; Y_1| U_{1c} , Q) \notag \\ & -I(U_{1a}; X_{2a}, X_{2b}| U_{1c}, Q)
+I(X_2, U_{1c}; Y_2|Q) \notag
\end{align}
for any distribution $p_{X_1 ,X_2, X_{2a}, X_{2b}, U_{1c}, U_{1a},Q}$.
For a given $p_{X_{2a}, X_{2b}, U_{1c}, U_{1a},Q}$ of \cite[Th.1]{MaricGoldsmithKramerShamai07Eu}
consider a related distribution $p_{X_{2a}', X_{2b}', U_{1c}', U_{1a}',Q'}$ such that
\begin{align*}
  &(U_{1c}', U_{1a}',Q')=(U_{1c}, U_{1a},Q)
\\&X_{2b}'=(X_{2a}, X_{2b}), \;\; X_{2a}'=\emptyset
\end{align*}
All rate constraints but~\eqref{eq:maric scheme: R1 1} are the same
under both distributions. Comparing~\eqref{eq:maric scheme: R1 1}
under the two distributions:
{\small \begin{align*}
&\rm{\eqref{eq:maric scheme: R1 1}}|_{p_{X_{2a}', X_{2b}', U_{1c}', U_{1a}',Q'}}\\
&  =I(U'_{1a}; Y_1| U'_{1c} , Q')-I(U_{1a}'; X_{2a}', X_{2b}'| U_{1c}', Q')+I(X_{2b}', U_{1c}'; Y_2 | X_{2a}', Q')\\
&  =I(U_{1a}; Y_1| U_{1c} , Q)-I(U_{1a}; X_{2a}, X_{2b}| U_{1c}, Q)+I(X_{2a},X_{2b}, U_{1c}; Y_2 | Q) \\
&  =I(U_{1a}; Y_1| U_{1c} , Q)-I(U_{1a}; X_{2a}, X_{2b}| U_{1c}, Q) +I(X_{2a}; Y_2| Q)+I(X_{2b}, U_{1c}; Y_2 | X_{2a}, Q) \\
& = I(X_{2a}; Y_2| Q)+{\rm{ \eqref{eq:maric scheme: R1 1}}}|_{p_{X_{2a}, X_{2b}, U_{1c}, U_{1a},Q}}\\
&  \geq \rm {\eqref{eq:maric scheme: R1 1}}|_{p_{X_{2a}, X_{2b}, U_{1c}, U_{1a},Q}}.
\end{align*}}

\bigskip
\noindent
\subsection{Containment of \cite[Thm. 1]{DevroyeThesis} in $\Rcal_{RTD}$}
\label{sec:devroye thesis achievable region}
We show this inclusion with the following steps:\\
\noindent
$\bullet$ We enlarge the region ${\cal R}_{DMT}$ by removing some rate constraints. \\
$\bullet$ We further enlarge the region by enlarging the set of possible input distributions. This allows us to remove the $V_{11}$  and $Q$ from the inner bound. We refer to this region as $\Rcal_{DMT}^{out}$  since is enlarges the original achievable region.\\
$\bullet$ We make a correspondence between the random variables and corresponding rates of $\Rcal_{DMT}^{out}$ and ${\cal R}_{RTD}$.  \\
$\bullet$ We choose a particular subset of ${\cal R}_{RTD}$, $\Rcal_{RTD}^{in}$, for which we can more easily show  $\Rcal_{DMT}\subseteq \Rcal_{DMT}^{out} \subset \Rcal_{RTD}^{in} \subseteq \Rcal_{RTD}$, since
 $\Rcal_{DMT}^{out}$ and $\Rcal_{RTD}^{in}$  have  identical input distribution decompositions and similar rate bound equations.

 {\bf Enlarge the region ${\cal R}_{DMT}$}\\
We first enlarge the rate region of \cite[Thm. 1]{DevroyeThesis}, $\Rcal_{DMT}$ by removing a number of constraints
(specifically, we remove equations (2.6, 2.8, 2.10, 2.13, 2.14, 2.16 2.17) of \cite[Thm. 1]{DevroyeThesis}) to obtain the region
$ \Rcal_{DMT}^{out} $ defined as the set of all rate pairs satisfying:
\begin{subequations}
\ea{
R_{21}' &=& I(V_{21}; V_{11}, V_{12}|W)
\label{eq: devroye 1}\\
R_{22}' &=& I(V_{22}; V_{11}, V_{12}|W)
\label{eq: devroye 2}\\
\nonumber \\
R_{11} &\leq&   I(Y_1,V_{12},V_{21}; V_{11}|W)
\label{eq: devroye 3}\\
R_{21}+R_{21}' &\leq& I(Y_1,V_{11},V_{12};V_{21}|W)
\label{eq: devroye 4}\\
R_{11}+R_{21}+R_{21}' &\leq& I(Y_1,V_{12};V_{11},V_{21}|W)+I(V_{11};V_{21}|W)
\label{eq: devroye 5}\\
R_{11}+R_{21}+R_{21}'+R_{12} &\leq& I(Y_1;V_{11},V_{21},V_{12}|W)+I(V_{11},V_{12};V_{21}|W)
\label{eq: devroye 6}\\
\nonumber \\
R_{22}+R_{22}' &\leq& I(Y_2,V_{12},V_{21}; V_{22}|W)
\label{eq: devroye 7}\\
R_{22}+R_{22}'+R_{21}+R_{21}' &\leq& I(Y_2,V_{12}; V_{22},V_{21}|W)+I(V_{22};V_{21}|W)
\label{eq: devroye 8}\\
R_{22}+R_{22}'+R_{21}+R_{21}' +R_{12} &\leq& I(Y_2; V_{22},V_{21},V_{12}|W)+I(V_{22},V_{21};V_{12}|W).
\label{eq: devroye 9}\\
}
\end{subequations}
taken over the union of  distributions
$$
p_{W}p_{V_{11}}p_{V_{12}} p_{X_1|V_{11},V_{12}} p_{V_{21}| V_{11} V_{12}} p_{V_{22}| V_{11},V_{12}} p_{X_2| V_{11},V_{12}, V_{21}, V_{22}}.
$$

Following the line of thoughts in  \cite[Appendix D]{willems1985discrete} it is possible to show that without loss of generality we can set $X_1$ to be a deterministic function of $V_{11}$ and $V_{12}$, allowing us insert $X_1$ next to $V_{11},V_{12}$ as follows:
\begin{subequations}
\ea{
R_{21}' &=& I(V_{21}; X_1, V_{11}, V_{12}|W)
\label{eq: devroye 1}\\
R_{22}' &=& I(V_{22}; X_1, V_{11}, V_{12}|W)
\label{eq: devroye 2}\\
\nonumber \\
R_{11} &\leq&   I(Y_1,V_{12},V_{21}; V_{11}|W)
\label{eq: devroye 3}\\
R_{21}+R_{21}' &\leq& I(Y_1,X_1,V_{11},V_{12};V_{21}|W)
\label{eq: devroye 4}\\
R_{11}+R_{21}+R_{21}' &\leq& I(Y_1,V_{12};V_{11},V_{21}|W)+I(V_{11};V_{21}|W)
\label{eq: devroye 5}\\
R_{11}+R_{21}+R_{21}'+R_{12} &\leq& I(Y_1;X_1,V_{11},V_{12},V_{21}|W)+I(X_1,V_{11},V_{12};V_{21}|W)
\label{eq: devroye 6}\\
\nonumber \\
R_{22}+R_{22}' &\leq& I(Y_2,V_{12},V_{21}; V_{22}|W)
\label{eq: devroye 7}\\
R_{22}+R_{22}'+R_{21}+R_{21}' &\leq& I(Y_2,V_{12}; V_{22},V_{21}|W)+I(V_{22};V_{21}|W)
\label{eq: devroye 8}\\
R_{22}+R_{22}'+R_{21}+R_{21}' +R_{12} &\leq& I(Y_2; V_{22},V_{21},V_{12}|W)+I(V_{22},V_{21};V_{12}|W)
\label{eq: devroye 9}
}
\end{subequations}

Using the factorization of the auxiliary RV's,  we may insert $X_1$ next to $V_{11}$  in equation \eqref{eq: devroye 6}. 

For equation (\ref{eq: devroye 3}):
\pp{
R_{11} &\leq&   I(Y_1,V_{12},V_{21}; V_{11}|W)\\
& = & I(Y_1,V_{21}; V_{11}|V_{12},W)+I(V_{12};V_{11}|W)\\
& = & I(Y_1,V_{21}; V_{11}|V_{12},W)\\
& = & I(Y_1,V_{21}; X_1,V_{11}|V_{12},W)\\
& = & I(Y_1;X_1, V_{11}| V_{12},V_{21},W)+I(V_{21}; X_1, V_{11} |V_{12},W).\\
}
For equation (\ref{eq: devroye 4}) we have:
\pp{
R_{11}+R_{21}+R_{21}' &\leq& I(Y_1,V_{12};V_{11},V_{21}|W)+I(V_{11};V_{21}|W)\\
& = & I(Y_1;V_{11},V_{21}|V_{12},W)+I(V_{12};V_{11},V_{21}|W)+I(V_{11};V_{21}|W)\\
& = & I(Y_1;V_{11},V_{21}|V_{12},W)+I(V_{12};V_{21}|V_{11},W)+I(V_{11};V_{21}|W)\\
& = & I(Y_1;V_{11},V_{21}|V_{12},W)+I(V_{11},V_{12};V_{21}|W)\\
& = & I(Y_1;X_1,V_{11},V_{21}|V_{12},W)+I(X_1,V_{11},V_{12};V_{21}|W)\\
}

The original region is thus equivalent to
\begin{subequations}
\ea{
R_{21}' &=& I(V_{21}; X_1,V_{11}, V_{12}|W)\\
R_{22}' &=& I(V_{22}; X_1,V_{11}, V_{12}|W)\\
\nonumber \\
R_{11} &\leq&  I(Y_1;X_1,V_{11}| V_{12},V_{21}|W)+I(V_{21}; X_1 |V_{12},W)\\
R_{21}+R_{21}' &\leq& I(Y_1,X_1,V_{11},V_{12};V_{21}|W)\\
R_{11}+R_{21}+R_{21}' &\leq&I(Y_1;X_1,V_{11},V_{21}|V_{12},W)+I(X_1;V_{21}|W)\\
R_{11}+R_{21}+R_{21}'+R_{12} &\leq& I(Y_1;X_1,V_{11},V_{21},V_{12}|W)+I(X_1,V_{11},V_{12};V_{21}|W)\\
\nonumber \\
R_{22}+R_{22}' &\leq& I(Y_2,V_{12},V_{21}; V_{22}|W)\\
R_{22}+R_{22}'+R_{21}+R_{21}' &\leq& I(Y_2,V_{12}; V_{22},V_{21}|W)+I(V_{22};V_{21}|W)\\
R_{22}+R_{22}'+R_{21}+R_{21}' +R_{12} &\leq& I(Y_2; V_{22},V_{21},V_{12}|W)+I(V_{22},V_{21};V_{12}|W)\\
}
\label{eq:devroye step 1}
\end{subequations}
taken over the union over all  distributions
$$
p_{W}p_{V_{11}}p_{V_{12}} p_{X_1|V_{11},V_{12}} p_{V_{21}| X_{1},V_{11} V_{12}} p_{V_{22}| X_{1},V_{11},V_{12}} p_{X_2| X_{1},V_{11},V_{12}, V_{21}, V_{22}}.
$$

\noindent
{\bf Enlarge the input distribution and eliminate $V_{11}$ and $W$ }\\
Now increase the set of possible input distribution of the input by letting $V_{11}$ to have any joint distribution with $V_{12}$. This is done by substituting $p_{V_{11}}$ with $p_{V_{11}| V_{12}}$ in the expression of the input distribution.  With this substitution we have: 
\pp{
p_{W}p_{V_{11}|V_{12}}p_{V_{12}} p_{X_1|V_{11},V_{12}} p_{V_{21}| X_{1},V_{11} V_{12}} p_{V_{22}| X_{1},V_{11},V_{12}} p_{X_2| X_1,V_{11},V_{12}, V_{21}, V_{22}}\\
\subseteq \; p_{W}p_{V_{12}} p_{V_{11},X_1|V_{12}} p_{V_{21}| X_{1},V_{11} V_{12}} p_{V_{22}| X_{1},V_{11},V_{12}} p_{X_2| X_1,V_{11},V_{12}, V_{21}, V_{22}}\\
= \;  p_{W}p_{V_{12}} p_{X_1'|V_{12}} p_{V_{21}| X_1', V_{12}} p_{V_{22}| X_1',V_{12}} p_{X_2| X_1',V_{12}, V_{21}, V_{22}}\\
}
with $X_1'=(X_1,V_{11})$.   Since $V_{12}$ is decoded at both decoders, the time sharing random $W$ may be incorporated with $V_{12}$ without loss of generality and thus can be dropped. The region described in (\ref{eq:devroye step 1}) is convex and time sharing does not increase the achievable region since the region is already convex.
With these simplifications, the region $\Rcal_{DMT}^{out}$ is now defined as
\begin{subequations}
\ea{
R_{21}' &=& I(V_{21}; X_1', V_{12})\\
R_{22}' &=& I(V_{22}; X_1', V_{12})\\
 \nonumber \\
R_{11} &\leq&  I(Y_1;X_1'| V_{12},V_{21})+I(V_{21}; X_1 |V_{12})\\
R_{21}+R_{21}' &\leq& I(Y_1,X_1',V_{12};V_{21})\\
R_{11}+R_{21}+R_{21}' &\leq&I(Y_1;X_1',V_{21}|V_{12})+I(X_1;V_{21})\\
R_{11}+R_{21}+R_{21}'+R_{12} &\leq& I(Y_1;X_1',V_{21},V_{12})+I(X_1',V_{12};V_{21})\\
 \nonumber  \\
R_{22}+R_{22}' &\leq& I(Y_2,V_{12},V_{21}; V_{22})\\
R_{22}+R_{22}'+R_{21}+R_{21}' &\leq& I(Y_2,V_{12}; V_{22},V_{21})+I(V_{22};V_{21})\\
R_{22}+R_{22}'+R_{21}+R_{21}' +R_{12} &\leq& I(Y_2; V_{22},V_{21},V_{12})+I(V_{22},V_{21};V_{12})
}
\end{subequations}
union over all the distributions
$$
p_{V_{12}} p_{X_1'|V_{12}} p_{V_{21}| X_1', V_{12}} p_{V_{22}| X_1',V_{12}} p_{X_2| X_1',V_{12}, V_{21}, V_{22}}\\
$$

 {\bf Correspondence between the random variables and rates.}
When referring to \cite{DevroyeThesis} please note that the index of the primary and cognitive user are reversed with respect to our notation (i.e $1 \goes 2$ and vice-versa). Consider the correspondences between the variables of \cite[Thm. 1]{DevroyeThesis} and those of Theorem \ref{thm:our achievable region} in Table \ref{tab:sec:devroye thesis achievable region} to obtain the region ${\cal R}_{DMT}^{out}$ defined as the set of rate pairs satisfying
\begin{table}
\centering
\begin{tabular}{| lll |} \hline
 RV, rate of Theorem \ref{thm:our achievable region}  &  RV, rate of  \cite[Thm. 1]{DevroyeThesis} & Comments \\ \hline
$U_{2c}, R_{2c}$ &$ V_{12}, R_{12}$ & TX 2 $\goes$ RX 1, RX 2\\  
$U_{1c}, R_{1c}$ & $V_{21}, R_{21}$ & TX 1 $\goes$ RX 1, RX 2\\  
$U_{1pb}, R_{1pb}$ & $V_{22}, R_{22}$ & TX 1 $\goes$ RX 1\\  
$X_2, R_{2pa}$ & $X_1', R_{11}$ & TX 2 $\goes$ RX 2 \\
$U_{2pb}=\emptyset, R_{2pb}'=0$ & -- &  TX 1 $\goes$  RX 2\\  
$R_{1c}' = I(U_{1c};X_2|U_{2c})$ & $L_{21}-R_{21} = I(V_{21};V_{11},V_{12})$ & Binning rate  \\ 
$R_{1pb}' = I(U_{1pb};X_2|U_{1c},U_{2c} )$ & $L_{22}-R_{22}  = I(V_{22};V_{11},V_{12})$ & Binning rates \\ 
$X_1$                 & $X_2$               &  \\
\hline
\end{tabular}
\caption{Assignment of RV's of Appendix  \ref{sec:devroye thesis achievable region} }\label{tab:sec:devroye thesis achievable region}
\end{table}

\begin{subequations}
\ea{
R_{1c}' &=& I(U_{1c}; X_2, U_{2c})
\label{e20}\\
R_{1pb}' &=& I(U_{1pb}; X_2, U_{2c})
\label{e21}\\
R_{2pa}+R_{1c}+R_{1c}'+R_{2c} &\leq& I(Y_2;U_{1c},U_{2c},X_2)+I(X_2,U_{2c};U_{1c})
\label{e23}\\
R_{2pa}+R_{1c}+R_{1c}' &\leq&I(Y_2;X_2,U_{1c}|U_{2c})+I(X_2;U_{1c})
\label{e24}\\
R_{1c}+R_{1c}' &\leq& I(Y_2,X_2,U_{2c};U_{1c})
\label{e25}\\
R_{2pa} &\leq&  I(Y_2;X_2| U_{2c},U_{1c})+I(U_{1c}; X_2 |U_{2c})
\label{e26}\\
R_{1pb}+R_{1pb}'+R_{1c}+R_{1c}' +R_{2c} &\leq& I(Y_1; U_{1pb},U_{1c},U_{2c})+I(U_{1pb},U_{1c};U_{2c})
\label{e27}\\
R_{1c}+R_{1pb}+R_{1c}'+R_{1pb}' &\leq& I(Y_1,U_{2c}; U_{1pb},U_{1c})+I(U_{1pb};U_{1c})
\label{e28}\\
R_{1pb}+R_{1pb}' &\leq& I(Y_1,U_{2c},U_{1c}; U_{1pb})
\label{e29}
}
\end{subequations}
taken over the union of all distributions
\ea{
p_{U_{2c}} p_{X_2|U_{2C}} p_{U_{1c}| X_2} p_{U_{1pb}| X_2} p_{X_1| X_2, U_{1c}, U_{1pb}}.
\label{eq:factorDevroye}
}

Next, we using the correspondences of the table and restrict the fully general input distribution of Theorem
 \ref{thm:our achievable region} to match the more constrained factorization of (\ref{eq:factorDevroye}), obtaining a region  $ \Rcal_{RTD}^{in} \subseteq \Rcal_{RTD}$ defined as the set of rate tuples satisfying
\begin{subequations}
\ea{
R_{1c}'&=&I(U_{1c}; X_2|U_{2c})
\label{e10}\\                                                     
R_{1c}'+R_{1pb}'&=& I(X_2; U_{1c}, U_{1pb}| U_{2c})
\label{e12}\\
R_{2c}+R_{1c}+R_{2pa}+R_{1c}' &\leq& I(Y_2; U_{2c},U_{1c},X_2)+I(U_{1c}; X_2 |U_{2c})
\label{e13}\\
R_{2pa}+R_{1c}+R_{1c}'&\leq& I(Y_2; U_{1c}, X_2| U_{2c})+I( U_{1c}; X_2| U_{2c})
\label{e14}\\
R_{1c}+R_{1c}' &\leq& I(Y_2; U_{1c}| U_{2c}, X_2)+I(U_{1c}; X_2| U_{2c})
\label{e15}\\
R_{2pa} &\leq&I(Y_2; X_2| U_{2c},U_{1c})+I(U_{1c}; X_2| U_{2c})
\label{e16}\\
R_{1pb}+R_{1pb}'+R_{1c}+R_{1c}' +R_{2c}  &\leq& I(Y_1; U_{2c},U_{1c},U_{1pb})
\label{e17}\\
R_{1c}+R_{1pb}+R_{1c}'+R_{1pb}' &\leq& I(Y_1; U_{1c},U_{1pb}|U_{2c})
\label{e18}\\
R_{1pb}+R_{1pb}' &\leq& I(Y_1; U_{1pb}|U_{2c},U_{1c})
\label{e19}
}
\end{subequations}
taken over the union of all distributions that factor as
$$
p_{U_{2c},X_2}  p_{U_{1c}| X_2 }p_{U_{1pb}| X_2} p_{X_1 | X_2, U_{1c}, U_{1pb}}.
$$


\noindent
{\bf Equation-by-equation comparison.} We now show that
$ \Rcal_{DMT}^{out}  \subseteq \Rcal_{RTD}^{in}$
by fixing an input distribution (which are the same for these two regions) and comparing the rate regions equation by equation. We refer to the equation numbers directly, and look at the difference between the corresponding equations in the two new regions.

\begin{itemize}
\item  \eqref{e13}-\eqref{e10} vs \eqref{e23}-\eqref{e20}: Noting the cancelation / interplay between the binning rates, we see that
$$
\lb \eqref{e13}-\eqref{e10}  \rb - \lb  \eqref{e24}-\eqref{e20} \rb=0.
$$
\item \eqref{e14}-\eqref{e10} vs. \eqref{e24}-\eqref{e20}: \pp{
\lb \eqref{e14}-\eqref{e10} \rb - \lb \eqref{e24}-\eqref{e20} \rb\\
\lag  \lag =-I(X_2;U_{1c})+I(U_{1c}; X_2,U_{2c}) \\
\lag  \lag =I(U_{2c}; U_{1c}|X_2)\\
\lag  \lag =0
}
\item \eqref{e15}-\eqref{e10} vs. \eqref{e25}-\eqref{e20}: again noting the cancelations, 
\pp{
\lb \eqref{e15}-\eqref{e10}\rb - \lb \eqref{e25}-\eqref{e20} \rb=0
}
\item \eqref{e16} vs. \eqref{e26}: 
\pp{
\eqref{e16} -\eqref{e26}=0 \\
}
\item  \eqref{e17}-\eqref{e12} vs. 
\eqref{e27}-\eqref{e21}-\eqref{e20}
\pp{
(\eqref{e17}-\eqref{e12})- (\eqref{e27}-\eqref{e21}-\eqref{e20})\\
\lag =-I(X_2; U_{1c}, U_{1pb}| U_{2c})\\
\lag  \lag   -I(U_{1pb},U_{1c}; U_{2c})+I(U_{1c}; U_{2c}, X_2)+I(U_{1pb}; U_{2c}, X_2)\\
\lag =-I(U_{1pb}, U_{1c};X_2, U_{2c}) +I(U_{1c}; U_{2c}, X_2)+I(U_{1pb}; U_{2c}, X_2)\\
\lag= -I(U_{1pb};X_2, U_{2c}) -I(U_{1c}; X_2, U_{2c}| U_{1pb})+I(U_{1c}; U_{2c}, X_2)+I(U_{1pb}; U_{2c}, X_2)\\
\lag= -I(U_{1c}; X_2, U_{2c}| U_{1pb})+I(U_{1c}; U_{2c}, X_2)\\
\lag=- H(U_{1c}| U_{1pb})+ H(U_{1c}| X_2, U_{2c},U_{1pb})+H(U_{1c})-H(U_{1c}| X_2, U_{2c})\\
\lag=I(U_{1c}; U_{1pb}) >0
}
where we have used the fact that $U_{1c}$ and $U_{1pb}$ are conditionally independent given $(U_{2c}, X_2)$.
\item  $\eqref{e18}-\eqref{e12}$ vs. $\eqref{e28}-\eqref{e21}-\eqref{e20}$:
\pp{
(\eqref{e18}-\eqref{e12})-(\eqref{e28}-\eqref{e21}-\eqref{e20})  \\
\lag = -I(X_2; U_{1c}, U_{1pb}|U_{2c})-I(U_{2c}; U_{1c}, U_{1pb})+I(U_{1pb};U_{2c}, X_2) - I(U_{1pb};U_{1c})+I(U_{1c}; X_2,U_{2c}) \\
\lag = -I(X_2,U_{2c}; U_{1c}, U_{1pb})+I(U_{1pb};U_{2c}, X_2) - I(U_{1pb};U_{1c})+I(U_{1c}; X_2,U_{2c}) \\
\lag = -I(X_2,U_{2c}; U_{1pb})-I(U_{1c}; X_2,U_{2c}| U_{1pb})+I(U_{1pb};U_{2c}, X_2) - I(U_{1pb};U_{1c})+I(U_{1c}; X_2,U_{2c}) \\
\lag = -I(U_{1c}; X_2,U_{2c},U_{1pb}) +I(U_{1c}; X_2,U_{2c}) \\
\lag = -I(U_{1c}; X_2,U_{2c})-I(U_{1c}; U_{1pb}|X_2,U_{2c}) +I(U_{1c}; X_2,U_{2c}) \\
\lag = 0
}
where we have used the fact that $U_{1c}$ and $U_{1pb}$ are conditionally independent given $(U_{2c}, X_2)$.
\item  $\eqref{e19}-\eqref{e12} +\eqref{e10}$  vs. $\eqref{e29}-\eqref{e21}$:
\pp{
(\eqref{e19}-\eqref{e12} +\eqref{e10}) - (\eqref{e29}-\eqref{e21}) \\
\lag = -I(U_{1pb}; X_2| U_{2c}, U_{1c})-I(U_{1pb}; U_{2c}, U_{1c})+I(U_{1pb}; X_2, U_{2c}) \\
\lag = -I(U_{1pb}; X_2, U_{2c}, U_{1c} )+I(U_{1pb}; U_{2c}, X_2 )\\
\lag =- I(U_{1pb}; U_{1c} | U_{2c}, X_2) \\
\lag= 0
}
\end{itemize}

\bigskip


\subsection{Containment of \cite[Thm. 2]{biao2009} in $\Rcal_{RTD}$}
\label{sec:biao}

The independently derived region in \cite[Thm. 2]{biao2008} uses a similar encoding structure as that of $\Rcal_{RTD}$ with two exceptions: a) the binning is done sequentially rather than jointly as in $\Rcal_{RTD}$ leading to binning constraints (43)--(45) in \cite[Thm. 2]{biao2008} as opposed to \eqref{eq:our achievable region R0'}--\eqref{eq:our achievable region R1'+R2'} in Thm.\ref{thm:our achievable region}. Notable is that both schemes have adopted a Marton-like binning scheme at the cognitive transmitter, as first introduced in the context of the CIFC in \cite{biao2008}. b) While the cognitive messages are rate-split in identical fashions, the primary message is split into 2 parts in \cite[Thm. 2]{biao2008} ($R_1=R_{11}+R_{10}$, note the reversal of indices) while we explicitly split the primary message into three parts $R_2 = R_{2c}+R_{2pa}+R_{2pb}$.
We show that the region of \cite[Thm.2]{biao2008}, denoted as ${\cal R}_{CC} \subseteq {\cal R}_{RTD}$ in  two steps:

\noindent $\bullet$  We first show that we may WLOG set $U_{11} = \emptyset$ in \cite[Thm.2]{biao2008}, creating a new region $R_{CC}'$.

\noindent $\bullet$  We next make a correspondence between our RV's and those of \cite[Thm.2]{biao2008} and  obtain identical regions. 

We note that the primary and cognitive indices are permuted in \cite{biao2008}.

We first show that $U_{11}$ in \cite[Thm. 2]{biao2008} may be dropped WLOG. Consider the  region ${\cal R}_{CC}$ of \cite[Thm. 2]{biao2008}, defined as the union over all distributions $p_{U_{10},U_{11},V_{11},V_{20},V_{22},X_1,X_2}p_{Y_1,Y_2|X_1,X_2}$ of all rate tuples satisfying:
\begin{align}
R_1 &\leq I(Y_{1};V_{11},U_{11}, V_{20}, U_{10})
\label{eq:CC 37}\\
R_2 &\leq I(Y_2;  V_{20}, V_{22}|U_{10}) -I(V_{22}, V_{20}; U_{11}| U_{10})
\label{eq:CC 38}\\
R_1+R_2 &\leq I(Y_1; V_{11}, U_{11}| V_{20}, U_{10})+I(Y_2; V_{22},V_{20},U_{10})- I(V_{22};U_{11},V_{11}|V_{20},U_{10})
\label{eq:CC 39}\\
R_1+R_2 & \leq I(Y_1; V_{11}, U_{11},V_{20}, U_{10})+I(Y_2; V_{22}|V_{20},U_{10})- I(V_{22};U_{11},V_{11}|V_{20},U_{10})
\label{eq:CC 40}\\
2 R_2+R_1 &\leq I(Y_1;V_{11},U_{11},V_{20}|U_{10})+I(Y_2; V_{22}|V_{20},U_{10}) + I(Y_2;V_{20},V_{22},U_{10})\notag \\& \;\;\;\;  - I(V_{22};U_{11},V_{11}|V_{20},U_{10})-I(V_{22}, V_{20}; U_{11}| U_{10})
\label{eq:CC 41}
\end{align}

Now let ${\cal R}_{CC}'$ be the region obtained by setting
$U_{11}'=\emptyset$ and $V_{11}'=(V_{11},U_{11})$
while keeping all remaining RV's identical. Then ${\cal R}_{CC}'$ is the union over all distributions $p_{U_{10},V'_{11},V_{20},V_{22},X_1,X_2}p_{Y_1,Y_2|X_1,X_2}$, 
with $V_{11}'=(V_{11},U_{11})$ in $\Rcal_{CC}$,
 of all rate tuples satisfying:
\begin{align}
R_1 &\leq I(Y_{1};V_{11},U_{11}, V_{20}, U_{10})
\label{eq:CC 37p}\\
R_2 &\leq I(Y_2;  V_{20}, V_{22}|U_{10})
\label{eq:CC 38p}\\
R_1+R_2 &\leq I(Y_1; V_{11}, U_{11}| V_{20}, U_{10})+I(Y_2; V_{22},V_{20},U_{10})- I(V_{22};U_{11},V_{11}|V_{20},U_{10})
\label{eq:CC 39p}\\
R_1+R_2 &\leq I(Y_1; V_{11}, U_{11},V_{20}, U_{10})+I(Y_2; V_{22}|V_{20},U_{10})- I(V_{22};U_{11},V_{11}|V_{20},U_{10})
\label{eq:CC 40p}\\
2 R_2+R_1 &\leq I(Y_1;V_{11},U_{11},V_{20}|U_{10})+I(Y_2; V_{22}|V_{20},U_{10})+ I(Y_2;V_{20},V_{22},U_{10}) \notag \\& \;\;\;\; - I(V_{22};U_{11},V_{11}|V_{20},U_{10})
\label{eq:CC 41p}
\end{align}
Comparing the two regions equation by equation, we see that
\begin{itemize}
  \item \eqref{eq:CC 37}= \eqref{eq:CC 37p}
  \item \eqref{eq:CC 38} $<$ \eqref{eq:CC 38p} as this choice of RV's sets the generally positive mutual information to 0
  \item \eqref{eq:CC 39}=\eqref{eq:CC 39p}
  \item \eqref{eq:CC 40}=\eqref{eq:CC 40p}
  \item \eqref{eq:CC 41} $<$ \eqref{eq:CC 41p} as this choice of RV's sets the generally positive mutual information to 0
\end{itemize}

\medskip

From the previous, we may set $U_{11} = \emptyset$ in  the region ${\cal R}_{CC}$ of \cite[Thm. 2]{biao2008} without loss of generality, obtaining the region ${\cal R}_{CC}'$ defined in \eqref{eq:CC 37p} -- \eqref{eq:CC 41p}.
We show that ${\cal R}_{CC}'$ may be obtained from the  region ${\cal R}_{RTD}$ with the assigment  of RV's, rates and binning rates
in Table \ref{tab:biao}. 


\begin{table}
\centering
\begin{tabular}{| lll |} \hline
 RV, rate of Theorem \ref{thm:our achievable region}  &  RV, rate of  \cite[Thm. 1]{DevroyeThesis} & Comments \\ \hline
$U_{2c}, R_{2c}$ &$ U_{10}, R_{10}$  & TX 2 $\goes$ RX 1, RX 2\\  
$X_{2}=U_{2c}$, $R_{2pa}=0$ & $U_{11}=\emptyset$, $R_{11}=0$      & TX 2 $\goes$ RX 2\\  
$U_{1c}, R_{1c}$ & $V_{20}, R_{20}$ & TX 1 $\goes$ RX 1, RX 2\\  
$U_{1pb}, R_{1pb}$ & $V_{22}, R_{22}$ & TX 1 $\goes$ RX 1\\  
$U_{2pb}, R_{2pb}$ &  $V_{11}$ &  TX 1 $\goes$  RX 2\\  
$ R_{1c}' $ &$L_{20}-R_{20}$ &\\
$ R_{1pb}' $&$L_{22}-R_{22}$ &\\
$ R_{2pb}' $&$L_{11}-R_{11}$ &\\
$X_1$                 & $X_2$               & \\
$X_2$                 & $X_1$               & \\
\hline
\end{tabular}
\caption{Assignment of RV's of Section \ref{sec:biao} }\label{tab:biao}
\end{table}


Evaluating  ${\cal R}_{CC}'$ defined by  \eqref{eq:CC 37p} -- \eqref{eq:CC 41p} with  the above assignment, translating all RV's into the notation used here, we obtain the region:
\pp{
R_{1c}' &\geq& 0 \\
R_{1pb}'+R_{2pb}'&\geq& I(U_{1pb}; U_{2pb} |U_{2c}, U_{1c}) \\
R_{2pb}+R_{2pb}' &\leq& I(Y_2; U_{2pb}|U_{2c}, U_{1c}) \\
R_{2pb}+R_{2pb}'+R_{1c}+R_{1c}' &\leq& I(Y_2; U_{1c}, U_{2pb}|U_{2c}) \\
R_{2pb}+R_{2pb}'+R_{1c}+R_{1c}'+R_{2c} &\leq& I(Y_2; U_{1c}, U_{2c}, U_{2pb}) \\
R_{1pb} +R_{1pb}' &\leq& I(Y_1;U_{1pb}| U_{2c}, U_{1c}) \\
R_{1pb}+R_{1pb}'+R_{1c}+R_{1c}' &\leq& I(Y_1;U_{1pb},U_{1c}| U_{2c}) \\
R_{1pb}+R_{1pb}'+R_{1c}+R_{1c}'+R_{2c} &\leq& I(Y_1;U_{1pb},U_{1c},U_{2c}) \\
}
Note that we may take binning rate equations $R_{1c}'\geq 0$ and $R_{1pb}'+R_{2pb}' \geq I(U_{1pb}; U_{2pb} |U_{2c}, U_{1c})$  to be equality without loss of generality - the largest region will take $R_{1c}', R_{1pb}', R_{2pb}'$ as small as possible.
The region ${\cal R}_{RTD}$ with $R_{2pa}=0$
\pp{
R_{1c}' &\geq& 0\\
R_{1c}'+R_{1pb}'&\geq& 0\\
R_{1c}'+R_{1pb}'+R_{2pb}' &\geq& I(U_{1pb}; U_{2pb} |U_{2c}, U_{1c}) \\
R_{2pb}+R_{2pb}' &\leq& I(Y_2;U_{2pb}|U_{2c}, U_{1c}) \\
R_{2pb}+R_{2pb}'+R_{1c}+R_{1c}' &\leq& I(Y_2; U_{1c}, U_{2pb}|U_{2c}) \\
R_{2pb}+R_{2pb}'+R_{1c}+R_{1c}'+R_{2c} &\leq& I(Y_2; U_{1c}, U_{2c}, U_{2pb}) \\
R_{1pb}+R_{1pb}' &\leq& I(Y_1;U_{1pb}| U_{2c}, U_{1c}) \\
R_{1pb}+R_{1pb}'+R_{1c}+R_{1c}' &\leq& I(Y_1;U_{1pb},U_{1c}| U_{2c}) \\
R_{1pb}+R_{1pb}'+R_{1c}+R_{1c}'+R_{2c} &\leq& I(Y_1;U_{1pb},U_{1c},U_{2c}) \\
}
For $R_{1c}'=0$ these two regions are identical, showing that ${\cal R}_{RTD}$ is surely no smaller than ${\cal R}_{CC}$.  For $R_{1c}'>0$, ${\cal R}_{RTD}$ , the binning rates of the region ${\cal R}_{RTD}$ are looser than the ones in ${\cal R}_{CC}$. This is probably due to the fact that the first one uses joint binning and latter one sequential binning.  Therefore ${\cal R}_{RTD}$ may produce rates larger than ${\cal R}_{CC}$. However, in general, no strict inclusion of ${\cal R}_{CC}$ in ${\cal R}_{RTD}$ has been shown.


\subsection{Containment of  \cite[Thm. 4.1]{jiang-achievable-BCCR} in ${\cal R}_{RTD}$:}
\label{sec:Jiang BCCR region}


In this scheme the common messages are created independently instead of having the common message from transmitter~1 being superposed to the common message from transmitter~2. The former choice introduces more rate constraints than the latter and allows us to show inclusion in $\Rcal_{RTD}$.


The region of \cite{jiang-achievable-BCCR} is expressed as  the set of all rate tuples satisfying
\begin{subequations}
\ea{
R_{22}' &\geq& I(W_2; V_1| U_1, U_2)
\label{Jiang 2}\\
R_{11}' + R_{22}' &\geq& I(W_2; W_1,V_1| U_1, U_2)
\label{Jiang 3}\\
R_{11}+R_{11}' &\leq& I(V_1, W_1; Y_1| U_1, U_2)
\label{Jiang 4}\\
R_{12}+R_{11}+R_{11}' &\leq& I(U_1, V_1 ,W_1 ; Y_1 | U_2)
\label{Jiang 5}\\
R_{21}+R_{11}+R_{11}' &\leq& I(U_2, V_1 ,W_1 ; Y_1 | U_1)
\label{Jiang 6}\\
R_{12}+R_{21}+R_{11}+R_{11}' &\leq& I(U_1, V_1, W_1, U_2; Y_1 )
\label{Jiang 7}\\
R_{22}+R_{22}' &\leq& I(W_2; Y_2 | U_1, U_2)
\label{Jiang 8}\\
R_{21}+R_{22}+R_{22}' &\leq& I(U_2,W_2; Y_2 | U_1)
\label{Jiang 9}\\
R_{12}+R_{22}+R_{22}' &\leq& I(U_1,W_2; Y_2 | U_2)
\label{Jiang 10}\\
R_{12}+R_{21}+R_{22}+R_{22}' &\leq& I(U_1,U_2,W_2; Y_2)
\label{Jiang 11}
}
\label{eq: Jiang-0}
\end{subequations}
taken over the union over of distributions
$$
p_{u_1}p_{v_1|u_1} p_{x_1|v_1,u_1}p_{u_2}p_{w_1,w_2| v_1,u_1, u_2} p_{x_0|w_1, w_2,v_1,u_1,u_2} p_{y_1,y_2|x_1,x_0}
$$
for
$(R_{11}',R_{22}',R_{11},R_{12},R_{21},R_{22})\in\RR^6_+.$

Following the argument of \cite[Appendix D]{willems1985discrete} we can show that  WLG we can take $X_1$ and $X_2$ to be deterministic functions, so that we can write
\begin{subequations}
\ea{
R_{22}' &\geq& I(W_2; V_1 ,X_1| U_1, U_2)
\label{Jiang 2}\\
R_{11}' + R_{22}' &\geq& I(W_2; W_1,V_1 ,X_1| U_1, U_2)
\label{Jiang 3}\\
R_{11}+R_{11}' &\leq& I(V_1,X_1, W_1; Y_1| U_1, U_2)
\label{Jiang 4}\\
R_{12}+R_{11}+R_{11}' &\leq& I(U_1, V_1, X_1 ,W_1 ; Y_1 | U_2)
\label{Jiang 5}\\
R_{21}+R_{11}+R_{11}' &\leq& I(U_2, V_1,X_1 ,W_1 ; Y_1 | U_1)
\label{Jiang 6}\\
R_{12}+R_{21}+R_{11}+R_{11}' &\leq& I(U_1, V_1, X_1 W_1, U_2; Y_1 )
\label{Jiang 7}\\
R_{22}+R_{22}' &\leq& I(W_2; Y_2 | U_1, U_2)
\label{Jiang 8}\\
R_{21}+R_{22}+R_{22}' &\leq& I(U_2,W_2; Y_2 | U_1)
\label{Jiang 9}\\
R_{12}+R_{22}+R_{22}' &\leq& I(U_1,W_2; Y_2 | U_2)
\label{Jiang 10}\\
R_{12}+R_{21}+R_{22}+R_{22}' &\leq& I(U_1,U_2,W_2; Y_2).
\label{Jiang 11}
}
\label{eq: Jiang-0}
\end{subequations}

We can now eliminate one random variable by  noticing that
\pp{
p_{u_1}p_{v_1|u_1} p_{x_1|v_1,u_1}p_{u_2}p_{w_1,w_2| v_1,u_1, u_2} p_{x_0|w_1, w_2,v_1,u_1,u_2} p_{y_1,y_2|x_1,x_0} \\
=p_{u_1} p_{v_1,x_1|u_1}p_{u_2}p_{w_1,w_2| v_1,u_1, x_1,u_2} p_{x_0|w_1, w_2,v_1,u_1,x_1,u_2} p_{y_1,y_2|x_1,x_0}
},
and setting  $V_1'=V_1,X_1$, to obtain the region

\begin{subequations}
\ea{
R_{22}' &\geq& I(W_2; V_1'| U_1, U_2)
\label{Jiang 2}\\
R_{11}' + R_{22}' &\geq& I(W_2; W_1,V_1'| U_1, U_2)
\label{Jiang 3}\\
R_{11}+R_{11}' &\leq& I(V_1', W_1; Y_1| U_1, U_2)
\label{Jiang 4}\\
R_{12}+R_{11}+R_{11}' &\leq& I(U_1, V_1' ,W_1 ; Y_1 | U_2)
\label{Jiang 5}\\
R_{21}+R_{11}+R_{11}' &\leq& I(U_2, V_1' ,W_1 ; Y_1 | U_1)
\label{Jiang 6}\\
R_{12}+R_{21}+R_{11}+R_{11}' &\leq& I(U_1, V_1' W_1, U_2; Y_1 )
\label{Jiang 7}\\
R_{22}+R_{22}' &\leq& I(W_2; Y_2 | U_1, U_2)
\label{Jiang 8}\\
R_{21}+R_{22}+R_{22}' &\leq& I(U_2,W_2; Y_2 | U_1)
\label{Jiang 9}\\
R_{12}+R_{22}+R_{22}' &\leq& I(U_1,W_2; Y_2 | U_2)
\label{Jiang 10}\\
R_{12}+R_{21}+R_{22}+R_{22}' &\leq& I(U_1,U_2,W_2; Y_2)
\label{Jiang 11}
}
\label{eq: Jiang-0}
\end{subequations}
taken over the union of all distributions of the form
$$
p_{u_1}p_{v_1'|u_1} p_{u_2}p_{w_1,w_2| v_1',u_1, u_2} p_{x_0|w_1, w_2,v_1',u_1,u_2} p_{y_1,y_2|v_1',x_0}
$$

We equate the RV's in the region of \cite{jiang-achievable-BCCR} with the RV's in Theorem \ref{thm:our achievable region} as in Table \ref{tab:Jiang BCCR region}.

\begin{table}
\centering
\begin{tabular}{| lll |} \hline
 RV, rate of Theorem \ref{thm:our achievable region}  &  RV, rate of  \cite[Thm. 1]{DevroyeThesis} & Comments \\ \hline
$U_{2c}, R_{2c}$ &$ U_{1}, R_{12}$  & TX 2 $\goes$ RX 1, RX 2\\  
$X_{2}, R_{2pa}$ & $V_{1}',R_{11}'$      & TX 2 $\goes$ RX 2\\  
$U_{1c}, R_{1c}$ & $U_{2}, R_{21}$ & TX 1 $\goes$ RX 1, RX 2\\  
$U_{1pb}, R_{1pb}$ & $W_{2}, R_{22}$ & TX 1 $\goes$ RX 1\\  
$U_{2pb}, R_{2pb}=0$ &  $W_{1}$ &  TX 1 $\goes$  RX 2\\  
$ R_{1c}' $ &$L_{20}-R_{20}$ &\\
$ R_{1pb}' $&$L_{11}-R_{11}$ &\\
$ R_{2pb}' $&$L_{22}-R_{22}$ &\\
$X_1$                 & $X_0$               & \\
$X_2$                 & $X_1$               & \\
\hline
\end{tabular}
\caption{Assignment of RV's of Section \ref{sec:Jiang BCCR region} }\label{tab:Jiang BCCR region}
\end{table}

With the substitution in the achievable rate region of  \reff{eq: Jiang-0}, we obtain the region
\begin{subequations}
\ea{
R_{1pb}' &\geq& I(U_{1pb}; X_2| U_{2c}, U_{1c})
\label{Jiang 1-0}\\
R_{1pb}' + R_{2pb}' &\geq &I(U_{1pb}; U_{2pb},X_2| U_{2c}, U_{1c})
\label{Jiang 1-1}\\
R_{2pa}+R_{2pb}' &\leq& I(X_2, U_{2pb}; Y_2| U_{2c}, U_{1c})
\label{Jiang 1-2}\\
R_{2c}+R_{2pa}+R_{2pb}' &\leq& I(U_{2c}, X_2 ,U_{2pb} ; Y_2 | U_{1c})
\label{Jiang 1-3}\\
R_{1c}+R_{2pa}+R_{2pb}' &\leq& I(U_{1c}, X_2 ,U_{2pb} ; Y_2 | U_{2c})
\label{Jiang 1-4}\\
R_{2c}+R_{1c}+R_{2pa}+R_{2pb}' &\leq& I(U_{2c}, X_2, U_{1c}, U_{1pb}; Y_2 )
\label{Jiang 1-5}\\
R_{1pb}+R_{1pb}' &\leq& I(U_{1pb}; Y_1 | U_{2c}, U_{1c})
\label{Jiang 1-6}\\
R_{1c}+R_{1pb}+R_{1pb}' &\leq& I(U_{1c},U_{1pb}; Y_1 | U_{2c})
\label{Jiang 1-7}\\
R_{2c}+R_{1pb}+R_{1pb}' &\leq& I(U_{2c},U_{1pb}; Y_1 | U_{1c})
\label{Jiang 1-8}\\
R_{2c}+R_{1c}+R_{1pb}+R_{1pb}' &\leq& I(U_{2c},U_{1c},U_{1pb}; Y_1)
\label{Jiang 1-9}
}
\label{eq: Jiang-1}
\end{subequations}
taken over the union of all distributions of the form
\begin{equation}
p_{U_{1c}}p_{U_{2c}}p_{X_2|U_{2c}} p_{U_{1pb}, U_{2pb}| U_{1c}, U_{2c}, X_2} p_{X_1| U_{2c}, U_{1c},U_{1pb},U_{2pb}}.
\label{eq:dist}
\end{equation}
%
%
%
%
%
%
Set  $R_{2pb}=0$ and $R_{1c}'=I(U_{1c}; X_2| U_{2c})$ in the achievable scheme of Theorem \ref{thm:our achievable region} 
and consider the factorization of the remaining RV's
%
%
as  \eqref{eq:dist}.
With this factorization of the distributions, we obtain the achievable region

\begin{subequations}
\ea{
R_{1c}'&=&I(U_{1c}; X_2| U_{2c})\\
{R_{1pb}'}&\geq&{I(U_{1pb};X_2|U_{2c},U_{1c})}\label{us 1 - 0}\\
R_{1pb}'+R_{2pb}' &\geq& I(U_{1pb}; X_2, U_{2pb}| U_{2c},U_{1c})
\label{us 1-1}\\
R_{2pa} +R_{2pb}' &\leq& I(Y_2; X_2,U_{2pb}| U_{2c},U_{1c})        +I(U_{1c}; X_2| U_{2c})
\label{us 1-2}\\
R_{1c}+R_{2pa}+R_{2pb}' &\leq& I(Y_2; U_{1c},       X_2,U_{2pb}| U_{2c})
\label{us 1-3}\\
 R_{2c}+R_{1c}+R_{2pa}+R_{2pb}' &\leq& I(Y_2; U_{2pb},U_{1c},U_{2c},X_2)
 \label{us 1-4}\\
R_{1pb}     +R_{1pb}' &\leq& I(Y_1;               U_{1pb}|U_{2c},U_{1c})
\label{us 1-5}\\
R_{1c}+R_{1pb}+R_{1pb}' &\leq& I(Y_1;        U_{1c},U_{1pb}|U_{2c})
\label{us 1-6}\\
 R_{2c}+R_{1c}+R_{1pb}+R_{1pb}' &\leq& I(Y_1; U_{2c},U_{1c},U_{1pb})
 \label{us 1-7}
}
\label{eq: us}
\end{subequations}
Note that with this particular factorization we have that $I(U_{1c}; X_2|U_{2c})=0$,  since $X_2$ is conditionally independent on $U_{1c}$ given {$U_{2c}$.}

We now compare the region of \reff{eq: Jiang-1} and \reff{eq: us} for a fixed input distribution, equation by equation:
\pp{
\reff{us 1 - 0}=\reff{Jiang 1-0}\\
\reff{us 1-1}=\reff{Jiang 1-1}\\
\reff{us 1-2}=\reff{Jiang 1-2}\\
\reff{us 1-3}=\reff{Jiang 1-4}\\
\reff{us 1-4}=\reff{Jiang 1-5}\\
\reff{us 1-5}=\reff{Jiang 1-6}\\
\reff{us 1-6}={\reff{Jiang 1-7}}\\
\reff{us 1-7}=\reff{Jiang 1-9}\\
}
clearly {\reff{Jiang 1-3}} and { \reff{Jiang 1-8}} are extra bounds that further restrict the region in \cite {jiang-achievable-BCCR} to be smaller than the region of Theorem \ref{thm:our achievable region}.

\end{appendix}


\end{document}